\documentclass[a4paper,UKenglish,cleveref, autoref, thm-restate]{lipics-v2021}

\hideLIPIcs  

\title{On Non-Deterministic Representation of Boolean Functions} 

\author{Sudharshan Rajagopalan}{Indian Institute of Technology Hyderabad, India}{co23btech11023@iith.ac.in}{}{}

\author{Nitin Saurabh}{Indian Institute of Technology Hyderabad, India}{nitin@cse.iith.ac.in}{https://orcid.org/0000-0002-1670-1114}{Supported by ANRF ARG-MATRICS grant ANRF/ARGM/2025/003191/MTR. He also acknowledges funding support from the National Quantum Mission (NQM), an initiative of the Department of Science and Technology (DST), Govt.~of India.}

\authorrunning{Sudharshan Rajagopalan and Nitin Saurabh} 

\Copyright{Sudharshan Rajagopalan and Nitin Saurabh} 



\category{} 

\relatedversion{} 

\acknowledgements{We would like to thank anonymous reviewers whose careful reading and thoughtful comments greatly improved the presentation of this paper. We would also like to highlight that no AI tools were used in this work. Nitin Saurabh also acknowledge the support provided by the Foundation for QC Innovation (FQCI), DST-NQM T-Hub at IISc Bengaluru, in facilitating the NQM project.}

\nolinenumbers 

\EventEditors{Raghavan Komondoor, Ramprasad Saptharishi, and James Worrell}
\EventNoEds{3}
\EventLongTitle{46th IARCS Annual Conference on Foundations of Software Technology and Theoretical Computer Science (FSTTCS 2026)}
\EventShortTitle{FSTTCS 2026}
\EventAcronym{FSTTCS}
\EventYear{2026}
\EventDate{December 16--18, 2026}
\EventLocation{IIT Delhi, New Delhi, India}
\EventLogo{}
\SeriesVolume{400}
\ArticleNo{42}

\hypersetup{colorlinks,linkcolor={blue!80!black},citecolor={green!50!black},urlcolor={red!40!black}}

\newtheorem{fact}[theorem]{Fact}

\newcommand{\zone}{\{0,1\}}

\newcommand{\sbool}{\{-1,1\}}

\newcommand{\reals}{\mathbb{R}}
\newcommand{\poly}{\mathsf{poly}}
\newcommand{\E}{\mathbb{E}}
\newcommand{\rank}{\mathsf{rank}}

\newcommand{\boolfn}[1]{\ensuremath {\zone}^{#1} \to \zone}

\newcommand{\ndeg}{\mathsf{ndeg}}
\newcommand{\rdeg}{\mathsf{rdeg}}
\newcommand{\fc}[2]{\ensuremath {\widehat{#1}(#2)}}

\newcommand{\dqc}{\mathsf{D}}

\newcommand{\AND}{\mathsf{AND}}

\begin{document}

\maketitle

\begin{abstract}
We study non-deterministic representation of Boolean functions by polynomials, introduced by (de Wolf, 2000). A polynomial $p(x)$ is said to be a \emph{non-deterministic polynomial} for a Boolean function $f$ if for all $a\in\zone^n$, $p(a)\neq 0$ iff $f(a)=1$. The \emph{non-deterministic degree} of $f$, denoted $\ndeg(f)$, is the minimal degree of a non-deterministic polynomial for $f$. In this work, we show the following main results:
\begin{alphaenumerate}
    \item Almost all Boolean functions on $n$ variables have non-deterministic degree at most $n/2 + O(\sqrt{n\log n})$. This confirms a conjecture of (Iyer et al., 2023). Together with the lower bound of (Iyer et al., 2023) it establishes a tight bound on the rational degree of a random Boolean function up to lower order terms. 
    \item We disprove a conjecture of (Kothari et al., 2026) about effective hypercube nullstellensatz for more than two polynomials. On the positive side, we also identify a constraint under which an effective hypercube nullstellensatz holds for many polynomials.   
\end{alphaenumerate}
\end{abstract}

\section{Introduction}
\label{sec:intro}

A Boolean function $f\colon\boolfn{n}$ is said to be \emph{exactly} represented by a polynomial $p(x_1,\ldots ,x_n)$ if for all $a\in\zone^n$, $p(a)=f(a)$. It is a well-known fact (see, e.g., \cite{BdW02}) that there exists a \emph{unique} multilinear poylnomial exactly representing a function $f$. The degree $\deg(f)$ of a Boolean function $f$ is defined to be the degree of this unique multilinear polynomial exactly representing $f$. Polynomials are probably the `nicest' object in mathematics which is attested not only by the existence of a mathematically rich theory but also by their ubiquitous presence in mathematics and sciences.  It is no wonder then that since the seminal work of Minsky and Papert \cite{MP69}, a long line of work in theoretical computer science has used various notions of representations by polynomials, e.g., \emph{sign representations}, \emph{pointwise approximations}, \emph{rational representations}, etc., to study different complexity measures associated with computing Boolean functions.   

In this work we focus on \emph{non-deterministic} representation of Boolean functions by polynomials. A polynomial $p(x_1,\ldots ,x_n)$ is said to be a \emph{non-deterministic polynomial} for a function $f\colon\boolfn{n}$ if $p(a)\neq 0$ iff $f(a)=1$. The \emph{non-deterministic degree} of $f$, denoted $\ndeg(f)$, is the minimum degree among all non-deterministic polynomials $p$ for $f$. This notion was introduced by de Wolf \cite{deWolf00} to study the \emph{non-deterministic quantum query complexity} of $f$. He showed that $\ndeg(f)$ equals the non-deterministic quantum query complexity of $f$ up to a constant factor. The non-deterministic degree is also related to another natural complexity measure of Boolean functions called the \emph{rational degree}. 

The \emph{rational degree} of $f$, denoted $\rdeg(f)$, is the minimum value of $\max\{\deg(p),\deg(q)\}$ such that $p$ and $q$ are polynomials satisfying $f(a)=p(a)/q(a)$ for all $a\in\zone^n$. It is a folklore result (credited to de Wolf; see \cite{IJKKDKSWW23,KKDWY26}) that $\rdeg(f)=\max\{\ndeg(f), \ndeg(\neg f)\}$. Mahadev and de Wolf \cite{MdW15} showed an even tighter connection between rational degree and quantum query complexity by observing that $\rdeg(f)$ exactly characterizes the exact postselected quantum query complexity (see also \cite{KKDWY26}). 

Long before the quantum connections of rational degree came to fore, the motivation to study rational degree was rooted in the classical complexity theory. It was observed by \cite{FFKL03} that if $\deg(f)$ is polynomially bounded by $\rdeg(f)$ then $\textup{P}=(\textup{C}_{=}\textup{P}\cap \textup{co-C}_{=}\textup{P})$ with respect to a generic oracle (see also \cite{F03}). Thus the question whether $\deg(f)$ is polynomially related to $\rdeg(f)$ was raised more than three decades ago in \cite{NS94}. 

In a recent breakthrough result \cite{KKDWY26} resolved the $\deg$ vs.~$\rdeg$ question by showing that $\dqc(f) = \tilde{O}(\ndeg(f)^{1.5}\ndeg(\neg f)^{1.5})$ where $\dqc(f)$ is the deterministic query complexity of $f$. They further observed that their main result gives an effective version of Hypercube Nullstellensatz. 
\begin{restatable}[{\cite[Theorem~10]{KKDWY26}}]{theorem}{NssRdeg}
    \label{thm:nss-rdeg}
    Let $g_1, g_2 \in \reals[x_1,\ldots ,x_n]$ such that they do not share any common zeroes on the hypercube $\zone^n$. Further suppose $g_1(a)\cdot g_2(a) = 0$ for all $a\in\zone^n$. Then there exist $h_1,h_2\in \reals[x_1,\ldots ,x_n]$ such that 
    \[h_1(a)g_1(a) + h_2(a)g_2(a) = 1\quad \text{ for all } a\in\zone^n,\] and \[\max\{\deg(\overline{h_1g_1}), \deg(\overline{h_2g_2})\} = O(\deg(g_1)^{1.5}\deg(g_2)^{1.5}\log\max\{\deg(g_1),\deg(g_2)\}),\]
    where the overline denotes multilinearization using the (Boolean) relations $x_i^2=x_i$ for all $i\in [n]$. 
\end{restatable}
They conjectured, in view of well-known classical nullstellensatz results (see \cite{jelonek05} and references therein), that the following natural generalization of \autoref{thm:nss-rdeg} to any number of polynomial holds. 
\begin{restatable}[{\cite[Conjecture~1]{KKDWY26}}]{conjecture}{ConjNssRdeg}
    \label{conj:nss-rdeg}
    For all integers $m\geq 3$, the following holds. Let $0 \neq g_1,\ldots ,g_m\in\reals[x_1,\ldots ,x_n]$. Suppose $g_1,\ldots ,g_m$ do not share any common zeroes on the hypercube $\zone^n$. Further suppose $g_1(a)\cdots g_m(a)=0$ for all $a\in\zone^n$. Then there exist $h_1,\ldots ,h_m\in\reals[x_1,\ldots ,x_n]$ such that 
    \[h_1(a)g_1(a)+\cdots  + h_m(a)g_m(a) = 1\quad  \text{ for all } a\in\zone^n,\] and 
    \[\max_{i\in[m]}\{\deg(\overline{h_ig_i})\} \leq \poly(\deg(g_1),\ldots ,\deg(g_m)).\]
\end{restatable}

In this work, we disprove \autoref{conj:nss-rdeg} (\autoref{thm:counter-nss-rdeg}) and prove a variant of it (\autoref{thm:nss-partition}). We also show that the counter-example in \autoref{thm:counter-nss-rdeg} is optimal (\autoref{thm:optimal-eg}) with respect to the number of polynomials and their degrees.  
They are presented in \autoref{sec:nss}. 

We also establish extremal bounds on the non-deterministic degree and rational degree of random Boolean functions. It was shown in \cite{IJKKDKSWW23} that all but a negligible fraction of Boolean functions have rational degree at least $n/2-O(\sqrt{n})$. They conjectured \cite[Conjecture~3]{IJKKDKSWW23} that an upper bound of $n/2+o(n)$ holds for a random Boolean function. We prove the conjecture (\autoref{thm:ndeg-random} and \autoref{cor:rdeg-random}), in \autoref{sec:ndeg-random}, by showing that all but a negligible fraction of Boolean functions have rational degree \emph{at most} $n/2+O(\sqrt{n\log n})$ and thereby establishing tight bounds up to lower order terms. 

We now present a brief overview of our results and techniques. 

\subsection{Our Results and Techniques}
In this paper, we first prove an extremal bound on non-deterministic degree of Boolean functions. 
\begin{restatable}{theorem}{NdegRandom}
    \label{thm:ndeg-random}
    All but a negligible fraction of Boolean functions on $n$ variables have non-deterministic degree at most $n/2+O(\sqrt{n\log n})$.
\end{restatable}

To establish the upper bound in the theorem, we prove the following technical lemma (\autoref{lem:ndeg-characterization}) that characterizes functions with $\ndeg(f) \leq d-1$. Let $H$ be the $2^n\times 2^n$ Hadamard matrix where the $(x,y)$-entry is given by $(-1)^{\langle x,y \rangle}$ (\autoref{defi:hadamard}). Let $H_d$ denote the submatrix of $H$ obtained by keeping the rows indexed by strings of Hamming weight at least $d$, and further let $H_d(f)$ denote the submatrix of $H_d$ obtained by keeping columns indexed by strings in $f^{-1}(1)$. 
\begin{restatable}{lemma}{NdegChar}
    \label{lem:ndeg-characterization}
    Let $f\colon\boolfn{n}$ be a Boolean function with $k:=|f^{-1}(1)|$. Then, 
    $\ndeg(f) \leq d-1$ if and only if there exists a coordinate-wise non-zero $v\in \reals^k$ such that $H_d(f)v=0$.  
\end{restatable}

To be able to fruitfully use the above lemma we further establish a linear algebraic characterization of matrices having a `coordinate-wise non-zero vector in the nullspace' (\autoref{lem:lin-alg-char}). It shows that for a coordinate-wise non-zero vector to exists in the nullspace, it is sufficient (and necessary) for the rowspace to have none of the standard $\zone$-unit vectors $e_j^t$. This in turn requires that rows remain independent even after removing a column $j$. That is, the matrix remains full rank even after removing a column.  

Observe that $H_d(f)$ is random submatrix of $H_d$ obtained by keeping each column with probability $1/2$. We need to show that $\rank(H_{d,-j}(f))$ is full with high probability, where $H_{d,-j}(f)$ denote the submatrix of $H_d(f)$ obtained by removing its $j$-th column. Now using the random Gram matrix $H_{d}(f)H_d(f)^t$ and the matrix Chernoff inequalities (\autoref{lem:matrix-chernoff}) we show that $H_d(f)$ satisfies the characterization of \autoref{lem:lin-alg-char} with very high probability. 

As a corollary to \autoref{thm:ndeg-random}, we also obtain the following extremal bound on rational degree proving a conjecture of \cite{IJKKDKSWW23}. 
\begin{restatable}{corollary}{RdegRandom}
    \label{cor:rdeg-random}
    All but a negligible fraction of Boolean functions on $n$ variables have rational degree at most $n/2 + O(\sqrt{n\log n})$.
\end{restatable}

Our second result is a counterexample to \autoref{conj:nss-rdeg} about effective hypercube nullstellensatz for many polynomials (made in \cite{KKDWY26}). 
\begin{restatable}{theorem}{CounterNssRdeg}
    \label{thm:counter-nss-rdeg}
    There exist three non-zero polynomials $g_1,g_2,g_3 \in \reals[x_1,\ldots ,x_n,y_1,\ldots ,y_n]$ with the following properties:
    \begin{romanenumerate}
        \item each of degree two, 
        \item do not share any common zeroes in $\zone^n\times \zone^n$, and 
        \item $g_1(a,b)\cdot g_2(a,b)\cdot g_3(a,b)=0$ for all $(a,b)\in\zone^n\times\zone^n$. 
    \end{romanenumerate}
    Let $p_1,p_2,p_3\in\reals[x_1,\ldots x_n,y_1,\ldots ,y_n]$ be any (multilinear) polynomials satisfying  
    \begin{itemize}
        \item $g_i(a,b) = 0 \implies p_i(a,b)=0$ for all $i\in\{1,2,3\}$ and $(a,b)\in\zone^n\times\zone^n$, and 
        \item $p_1(a,b) + p_2(a,b) + p_3(a,b) =1$ for all $(a,b)\in\zone^n\times\zone^n$.
    \end{itemize}
    Then,  
    \[\max\{ \deg(p_1), \deg(p_2), \deg(p_3)\} \geq n.\]
\end{restatable}

This counterexample is optimal with respect to number of polynomials and their degrees as we state below. 
\begin{restatable}{theorem}{OptimalEg}
    \label{thm:optimal-eg}
    There exists \emph{no} set of $3$ degree-$1$ polynomials that is a counterexample to \autoref{conj:nss-rdeg}. 
    
    In other words, let $g_1,g_2,g_3\in\reals[x_1,\ldots ,x_n]$ be three non-zero affine-linear forms such that they do not share any common zeroes on the hypercube $\zone^n$. Further, suppose $g_1(a)g_2(a) g_3(a)=0$ for all $a\in\zone^n$. Then, there exist $h_1,h_2 ,h_3\in\reals[x_1,\ldots ,x_n]$ such that 
    \[h_1(a)g_1(a)+ h_2(a)g_2(a) + h_3(a)g_3(a) = 1\quad  \text{ for all } a\in\zone^n,\] and 
    \[\max\{\deg(h_1), \deg(h_2), \deg(h_3)\} = O(1).\] 
\end{restatable}

We also give a counterexample comprising of $5$ linear polynomials (\autoref{prop:counter-nss-rdeg-linear}). These counterexamples build on a similar example (\autoref{prop:counter-eg-rdeg-partial}) from \cite{KKDWY26}.  
Finally, we identify an extra constraint on the zero set of polynomials that allows us to prove an effective version of hypercube nullstellensatz for many polynomials. 
\begin{restatable}{theorem}{NssPartition}
    \label{thm:nss-partition}
    Let $m \geq 3$ be an integer and $g_1, \ldots , g_m \in \reals[x_1, \ldots ,x_n]$ be non-zero polynomials satisfying the following:
    \begin{romanenumerate}
        \item they do not share any common zeroes in $\zone^n$, 
        \item $g_1(a)\cdots g_m(a) = 0$ for all $a\in\zone^n$, and 
        \item there exists a non-empty $S \subset [m]$ such that 
        \(\left(\cup_{i \in S}V_i\right)\cap \left(\cup_{i\notin S}V_i\right) = \emptyset\), where $V_i := g_i^{-1}(0) \subseteq \zone^n$ for $i\in [m]$. 
    \end{romanenumerate}
    Then, there exist (multilinear) polynomials $p_1,\ldots ,p_m \in \reals[x_1,\ldots ,x_n]$ such that the following holds:
    \begin{alphaenumerate}
        \item $g_i(a) = 0 \implies p_i(a)=0$ for all $i\in[m]$ and $a\in\zone^n$,
        \item $p_1(a)+\cdots  + p_m(a)=1$ for all $a\in\zone^n$, and 
        \item \(\max_{i\in [m]}\{\deg(p_i)\} \leq \poly(\deg(g_1),\ldots ,\deg(g_m) )\). 
    \end{alphaenumerate}
\end{restatable}

\section{Preliminaries}
\label{sec:prelims}

For a positive integer $n$, we denote the set $\{1,\ldots ,n\}$ by $[n]$. We use $\sqcup$ to denote the disjoint union of sets. A polynomial $p(x_1,\ldots ,x_n) \in \reals[x_1,\ldots ,x_n]$ is said to be multilinear if it is of the form $\sum_{S \subseteq [n]}c_S\cdot \prod_{i\in S}x_i$ with $c_S\in\reals$. It is a well-known fact (see, e.g., \cite{BdW02}) that every $f\colon\zone^n\to\reals$ is exactly represented by a unique multilinear polynomial $p(x_1,\ldots ,x_n)$.  
\begin{definition}
    \label{defi:degree} 
    The \emph{degree} of $f\colon\boolfn{n}$, denoted $\deg(f)$, is the degree of the unique multilinear polynomial $p(x)$ exactly representing $f$, i.e., $p(a)=f(a)$ for all $a\in\zone^n$. 
\end{definition}

We will use Fourier representation of Boolean functions. In particular, consider the space of all functions from $\zone^n$ to $\reals$ with the following inner-product  
\begin{align*}
    \langle f, g\rangle := \E_{x\in\zone^n}[f(x)g(x)], 
\end{align*}
where the expectation is with respect to the uniform distribution on $\zone^n$. For $S\subseteq [n]$, define the character functions $\chi_S\colon\zone^n\to\sbool$ as $\chi_S(x)=(-1)^{\sum_{i\in S}x_i}$. Then the set of character functions $\{\chi_S\}_{S\subseteq [n]}$ forms an orthonormal basis for the space of all real-valued functions on $\zone^n$. Hence, every real-valued function $f\colon\zone^n\to\reals$ has a unique Fourier expansion 
\begin{align*}
    f(x) = \sum_{S\subseteq [n]} \fc{f}{S}\chi_S(x), \text{ where the Fourier coefficients }\fc{f}{S}\in\reals .  
\end{align*}
It is easily seen that the Fourier-degree $\max\{|S|\mid \fc{f}{S}\neq 0\}$ equals $\deg(f)$. 
From the orthonormality, it follows that $\fc{f}{S}=\E_{x\in\zone^n}[f(x)\chi_S(x)]$. Fourier representation is intimately tied with the Hadamard matrices. 
\begin{definition}
    \label{defi:hadamard}
    Let $n$ be an integer. Then, the Hadamard matrix $H$ of dimension $2^n \times 2^n$ is defined such that the rows and columns are indexed by Boolean strings in $\zone^n$ and $(x,y)$-th entry of $H$ equals $(-1)^{\sum_{i=1}^nx_iy_i}$. 
\end{definition}
It is then easily seen that for a function $f\in\reals^{2^{n}}$, $\hat{f}=\frac{1}{2^n}Hf$,  where the vector of Fourier coefficients $\hat{f}\in\reals^{2^n}$. Furthermore, $f=H\hat{f}$.

\begin{definition}
    A polynomial $p(x)$ is said to be a \emph{non-deterministic polynomial} for $f\colon\boolfn{n}$ if for all $a\in\zone^n$, $p(a)\neq 0$ if and only if $f(a)=1$. The \emph{non-deterministic degree} of $f$, denoted $\ndeg(f)$, is the minimum degree among all non-deterministic polynomials for $f$. 
\end{definition}

\begin{definition}
    Polynomials $p(x)$ and $q(x)$ are said to be a \emph{rational polynomial} for $f\colon\boolfn{n}$ if for all $a\in\zone^n$, $q(a)\neq 0$ and $p(a)/q(a) = f(a)$. The \emph{rational degree} of $f$, denoted $\rdeg(f)$, is the minimum of $\max\{\deg(p),\deg(q)\}$ among all rational polynomials for $f$. 
\end{definition}

We note the following (folklore) fact about rational degree and non-deterministic degree. For a proof see \cite{KKDWY26}. 

\begin{fact}
\label{fact:rdeg-ndeg}
For every Boolean function $f$, $\rdeg(f) = \max\{\ndeg(f),\ndeg(\neg f)\}$. 
\end{fact}

We note the following facts from linear algebra.

\begin{fact}
    Let $p(x_1,\ldots ,x_n)\in \reals[x_1,\ldots ,x_n]$. Then $p=0$ in $\reals[x_1,\ldots ,x_n]$ if and only if $p\colon\reals^n\to\reals$ is the zero function. 
\end{fact}

We need the following fact about the non-zeroness of a multilinear polynomial. For a proof see, e.g., \cite{NS94}. 

\begin{lemma}[Ore-Schwartz-Zippel-DeMillo-Lipton Lemma]
    \label{lem:schwartz-zippel}
    Let $p(x_1,\ldots ,x_n)\in \reals[x_1,\ldots ,x_n]$ be a non-zero multilinear polynomial of degree $d$. Then, $\Pr_x[p(x)\neq 0] \geq 2^{-d}$, where $x$ is chosen uniformly at random from $\zone^n$.  
\end{lemma}
We also note the following anti-concentration result about linear forms \cite{Erdos45}. 
\begin{lemma}[\cite{Erdos45}]
    \label{lem:littlewood-offord}
    Let $a_1,\ldots ,a_n$ be $n$ real numbers with $|a_i| \geq 1$. Further let $\epsilon_i \in \sbool$ for each $i\in[n]$. Then the number of sums $\sum_{i=1}^n\epsilon_ia_i$ which fall in the interior of an arbitrary interval $I$ of length $2$ is at most $\binom{n}{\lfloor n/2\rfloor}$.  
\end{lemma}

Let $e_j \in \reals^n$ denote the $\zone$-unit-vector with $1$ at the $j$-th location and $0$ everywhere else. For a vector $v\in\reals^n$, let $v^{t}$ denote its transpose. 
\begin{definition}
    Let $A$ be a $m\times n$ matrix over $\reals$. Then, the \emph{nullspace} of $A$ is defined to be the set of all vectors $v\in \reals^n$ such that $Av=0$. It is a subspace of $\reals^n$. 

    Further, the \emph{rowspace} of $A$ is defined to be the subspace in $\reals^n$ generated by the rows of $A$. That is, it is the set of all vectors $v^{t}A$ where $v\in\reals^m$. 
\end{definition}

\begin{fact}
\label{fact:rowspace-nullspace}
    Let $A$ be a $m\times n$ matrix over $\reals$. Then, the rowspace of $A$ is the \emph{orthogonal complement} of the nullspace of $A$. That is, $(\textup{rowspace of } A)=(\textup{nullspace of }A)^{\perp}$. 
\end{fact}

We need the following version of Chernoff-Hoeffding concentration inequality. 
    
\begin{lemma}[Chernoff-Hoeffding]
    \label{lem:chernoff-binary}
    Let $Y_1,\ldots ,Y_n$ be independent and identically distributed $\zone$-random variables with $\Pr(Y_i=1)=\Pr(Y_i=0)=1/2$. Let $Y=\sum_{i=1}^nY_i$. Then, for any $t>0$, \[\Pr(Y\geq \E[Y]+t) \leq e^{-2t^2/n}.\]
\end{lemma}
We also need the following matrix Chernoff inequalities from \cite{Tropp12,Tropp15}. See \cite[Corollary~5.2]{Tropp12} or \cite[Theorem~5.1.1]{Tropp15} for the version stated here. See also Remark~$5.3$ in \cite{Tropp12}. 
\begin{lemma}[Matrix Chernoff Lower Tail Inequality {\cite{Tropp12,Tropp15}}]
    \label{lem:matrix-chernoff}
    Consider a finite sequence $\{Y_k\}$ of independent, random, self-adjoint matrices with common dimension $d$ that satisfy 
    \[Y_k \succcurlyeq 0 \text{ and } \lambda_{\max}(Y_k) \leq L \text{ for each index }k.\]
    Define $Y:=\sum_kY_k$ and the minimum eigenvalue $\mu_{\min}$ of the expectation $\E[Y]$ as $\mu_{\min} := \lambda_{\min}(\E[Y])$. Then, for $\delta \in [0,1]$, 
    \[\Pr\left(\lambda_{\min}(Y) \leq (1-\delta)\mu_{\min} \right) \leq d \cdot \left[\frac{e^{-\delta}}{(1-\delta)^{1-\delta}}\right]^{\mu_{\min}/L} \leq d\cdot e^{-\delta^2\mu_{\min}/2L}.\]
\end{lemma}

\section{Extremal properties of non-deterministic polynomials}
\label{sec:ndeg-random}
In this section we give extremal bounds on non-deterministic polynomial representations of Boolean functions. We begin with a simple characterization of functions having full non-deterministic degree. It is well-known that non-deterministic degree of $\AND_n$ is $n$ \cite{AF93,deWolf00}. (In fact, in combinatorics it is known as the Alon-F\"{u}redi theorem.) We observe that only $\AND_n$ have non-deterministic degree $n$. 

\begin{definition}
    \label{defi:indicator-function}
    A Boolean function $f\colon\boolfn{n}$ is called an \emph{indicator function} if $|f^{-1}(1)|=1$. 
\end{definition}
\begin{proposition}
    \label{prop:ndeg-equals-n}
    Let $f\colon\boolfn{n}$ be a Boolean function. Then, $\ndeg(f)=n$ if and only if $f$ is an indicator function. 
\end{proposition}
\begin{proof}
    As mentioned before, the Alon-F\"{u}redi theorem \cite{AF93} proves that the non-deterministic degree of an indicator function is $n$ (see also \cite{deWolf00}). What remains to be shown is that if $\ndeg(f)=n$ then $f$ must be an indicator function. To establish this, we prove the contrapositive. That is, if $f$ is \emph{not} an indicator function then $\ndeg(f)\leq n-1$.

    Let $f$ be a Boolean function such that $|f^{-1}(1)|:=k \geq 2$, and $a_1,\ldots ,a_k$ be the $k$-elements of $f^{-1}(1)$. For $i\in[k]$, define $I_i$ to be the indicator function of the input $a_i$, i.e., $I_i(x)=1$ iff $x=a_i$. Recall, $\deg(I_i) =n$. Let $c_i$ be the non-zero coefficient of the degree $n$ monomial $\prod_{j=1}^n x_j$ in $I_i$. Define $\lambda := -\sum_{i=1}^{k-1}c_i^2/c_k$. Clearly $\lambda$ is well-defined and $\neq 0$. Now consider the polynomial \[p(x) = \sum_{i=1}^{k-1} c_iI_i(x) + \lambda I_k(x).\] 
    We now claim that $p(x)$ is a non-deterministic polynomial for $f$ of degree at most $n-1$. It is easily seen from construction that $p(x)$ evaluates to non-zero iff $x\in\{a_1,\ldots ,a_k\}$. Moreover, the coefficient of the degree $n$ monomial equals $\sum_{i=1}^{k-1}c_i^2+\lambda c_k$ which equals $0$ by the choice of $\lambda$. 
\end{proof}

The proof, in \autoref{prop:ndeg-equals-n}, showing that non-indicator functions have non-deterministic degree at most $n-1$ is simple, and yet instructive enough to allow us to obtain a characterization of non-deterministic degree of arbitrary Boolean functions. This characterization in turn will help us bound the non-deterministic degree of random Boolean functions. Before we proceed further we recall the definition of Hadamard matrices (\autoref{defi:hadamard}). We need the following submatrices of them. 

\begin{definition}
    \label{defi:hadamard-restricted}
    For $0\leq d \leq n$ and $S\subseteq \zone^n $, let $H_d(S)$ denote the submatrix of $H$ obtained by keeping columns indexed by $S$ and rows indexed by strings of Hamming weight $\geq d$. When $S=f^{-1}(1)$ for a Boolean function $f$, then we will use the shorthand $H_d(f)$ to denote the submatrix $H_d(f^{-1}(1))$. When $S=\zone^n$, we denote the submatrix by $H_d$. 
\end{definition}
We are now ready to state the characterization. 
\NdegChar*
\begin{proof}
    Let $f^{-1}(1)=\{a_1,\ldots ,a_k\}\subseteq\zone^n$ and assume, without loss of generality, the columns of $H_d(f)$ are indexed from left to right by $a_1,\ldots , a_k$.
    
    ($\Rightarrow$) Let $\ndeg(f)\leq d-1$ and $p(x)$ be a non-deterministic polynomial for $f$ with degree at most $d-1$. Then define $v = (v_1,\ldots ,v_k)^{t} \in \reals^k$ as follows: $v_i = p(a_i)$ for all $i\in[k]$. Clearly $v$ is coordinate-wise non-zero. Further we claim $H_d(f)v=0$. To see this first recall that $\hat{p} = \frac{1}{2^n}Hp$. Secondly observe that $H_d$ is the restriction of $H$ to Fourier characters of degree at least $d$. Now since $\deg(p)\leq d-1$, we must have $0=H_{d}p = H_d(f)v$. 
    
    ($\Leftarrow$) Let $v\in\reals^k$ be coordinate-wise non-zero such that $H_d(f)v=0$. Define the function $p\colon\zone^n\to\reals$ as follows:
    \begin{align*}
        p(x) = \begin{cases} v_j, & \text{ if }x=a_j \in f^{-1}(1), \\ 0, & \text{otherwise}. \end{cases}
    \end{align*}    
    Clearly $p$ is a non-deterministic polynomial for $f$, i.e., $p(x) \neq 0 \iff x \in f^{-1}(1)$.  Therefore, $\ndeg(f)\leq \deg(p)$. We now claim that $\deg(p)\leq d-1$; thereby completing the proof. 

    To see that $\deg(p)\leq d-1$, consider the Fourier expansion $\hat{p}=\frac{1}{2^n}Hp$. Further consider the restriction $\hat{p}_d$ of $\hat{p}$ to Fourier characters of degree at least $d$. We have $\hat{p}_d = \frac{1}{2^n}H_dp = \frac{1}{2^n}H_d(f)v =0$. Thus $\deg(p)\leq d-1$. 
\end{proof}
We now prove a linear algebraic lemma that helps utilize the characterization in \autoref{lem:ndeg-characterization}. 
\begin{lemma}
    \label{lem:lin-alg-char}
    Let $A$ be a real matrix with $k$ columns. Then, there exists a coordinate-wise non-zero $v \in\reals^k$ in the nullspace of $A$ (i.e., $Av=0$) if and only if for all $j\in[k]$, $e_j^t$ is not in the rowspace of $A$.  
\end{lemma}
\begin{proof}
    It is a well-known fact in linear algebra that the rowspace of $A$ is the orthogonal complement of the nullspace of $A$ (\autoref{fact:rowspace-nullspace}). 

    ($\Rightarrow$) Let $v$ be a coordinate-wise non-zero vector in the nullspace of $A$. Then, since rowspace of $A$ is the orthogonal complement of the nullspace and $e_j^tv= v_j \neq 0$, we must have $e_j^t \notin$ rowspace of $A$ for all $j\in[k]$. 

    ($\Leftarrow$) Let $e_j^t$ is not in the rowspace of $A$ for all $j\in[k]$. By \autoref{fact:rowspace-nullspace}, for each $j\in[k]$, there must exist vectors $v^{(j)}$ in the nullspace of $A$ such that its $j$-th coordinate $v^{(j)}_j \neq 0$. We now claim that there is a linear combination of $v^{(j)}$'s that is coordinate-wise non-zero. 
    
    Let $v=\sum_{j=1}^k x_j v^{(j)}$ be an arbitrary linear combination of $v^{(j)}$'s. We need to show that there exist a choice of $x_j$'s such that $v$ is coordinate-wise non-zero. Let us consider each coordinate $v_i(x) = \sum_{j=1}^k x_jv_i^{(j)}$ as a polynomial in $x$. Note that, since $v_i^{(i)}\neq 0$, each $v_i(x)$ is a non-zero polynomial.  
    Now consider the product polynomial $p(x)=\prod_{i=1}^kv_i(x)$. Since each $v_i(x)$ is a non-zero polynomial, the product polynomial is also a non-zero polynomial (of degree $k$). Therefore, since $\reals$ is an infinite field, there exists a point $c\in \reals^k$ such that $p(c) \neq 0$, which in turn implies that each $v_i(c) \neq 0$, thereby proving the existence of a coordinate-wise non-zero $v$ in the nullspace of $A$.     
\end{proof}

We are now ready to prove that almost all Boolean functions have degree at most $n/2+O(\sqrt{n\log n})$. 

\NdegRandom* 
\begin{proof}
    Let us fix $d= n/2+\sqrt{cn\ln n}$, for an appropriate choice of the constant $c$ to be made later. Consider $H_d(f)$ for a random Boolean function $f$. By \autoref{lem:ndeg-characterization}, it suffices to show that with probability $1-o(1)$ there exists a coordinate-wise non-zero vector $v$ such that $H_d(f)v=0$. Which in turn, by \autoref{lem:lin-alg-char}, is equivalent to showing that for a uniformly at random $f$, with probability $1-o(1)$, no standard basis vectors $e_j^t$ are in the rowspace of $H_d(f)$. We now prove this. 

    Recall $H_d$ denote the submatrix of $H$ obtained by keeping those rows indexed by Boolean strings of Hamming weight $\geq d$. Let $R$ denote the number of rows in $H_d$. Then, $R=\sum_{j=d}^n\binom{n}{j}\leq 2^n\cdot e^{-2c\ln n}=2^n\cdot n^{-2c}$, where the inequality follows from the choice of $d$ and the Chernoff-Hoeffding inequality (\autoref{lem:chernoff-binary}).  

    Let us fix a Boolean function $f$. Suppose that a standard basis vector $e_j^t$ is in the row span of $H_d(f)$. Then there exists a non-zero vector $u\in\reals^R$ such that $u^tH_d(f)=e_j^t$. Let $H_{d,-j}(f)$ denote the submatrix of $H_d(f)$ obtained by removing its $j$-th column. Then, $u^tH_{d,-j}(f)=0$. That is, the rows of $H_{d,-j}(f)$ are linearly dependent. Therefore, if a standard basis vector $e_j^t$ is in the row span of $H_d(f)$ then the rows of $H_{d,-j}(f)$ are linearly dependent, and therefore $\rank(H_{d,-j}(f)) < R$. Consider the $R \times R$ Gram matrix $G^f_{j}:=H_{d,-j}(f)H_{d,-j}(f)^t$. Since $\rank(G^f_j)=\rank(H_{d,-j}(f))$, we must therefore have $\rank(G^f_j) < R$. 
    
    Hence, for a random $f$, \[\Pr\left(e_j^t \text{ is in the rowspace of }H_d(f)\right) \leq \Pr\left(\rank(H_{d,-j}(f))< R\right) = \Pr\left(\rank(G^f_j)< R\right).\] 

    We now estimate the probability that $\rank(G^f_j)< R$. Consider the matrix-valued function 
    \begin{align}
        \label{eq:defn-gram}
        G_j(\delta_i) = \sum_{\substack{i=1 \\ i\neq j}}^{2^n}\delta_iv_iv_i^t,
    \end{align}
    where $v_i \in \reals^R$ are the columns of $H_d$ and $\delta_i$'s are independent $\zone$-indicator variables. For a random $f$, $\delta_i\sim\text{Bernoulli}(1/2)$. Therefore,
    \begin{align}
        \label{eq:to-mat-chernoff}
        \Pr_f(\rank(G_j^f)< R) = \Pr_f(\lambda_{\min}(G_j(f)) \leq 0). 
    \end{align}
    We now bound the probability that minimum eigenvalue of the random Gram matrix $G_j(f)$ equals zero using the matrix Chernoff inequality (\autoref{lem:matrix-chernoff}). We note that the maximum eigenvalue $\lambda_{\max}(v_iv_i^t)$ of the rank-1 matrix $v_iv_i^t$ is $R$, since $v_i$ is a $\{1,-1\}$-vector of dimension $R$. We also compute the expected Gram matrix 
    \begin{align}
        \label{eq:exp-gram}
        \E_{f}[G_j(f)] = \frac{1}{2}H_{d,-j}H_{d,-j}^t = \frac{1}{2}\left(H_dH_d^t -v_jv_j^t\right) = \frac{1}{2}\left(2^nI_R -v_jv_j^t\right),
    \end{align}
    where $I_R$ is the $R\times R$ identity matrix and, recall $H_{d,-j}$ is the submatrix of $H_d$ obtained by removing its $j$-th column. 
    The first equality in \autoref{eq:exp-gram} follows from the linearity of expectation and writing the sum of outer products as matrix multiplication, the second follows from adding and subtracting the term $v_jv_j^t$, and finally the third follows from the fact that the rows of $H_d$ (Hadamard matrices) are mutually orthogonal. 
    
    It then follows from \autoref{eq:exp-gram} that the minimum eigenvalue of the expected Gram matrix $\mu_{\min}(\E[G_j])=(2^n-R)/2$. Therefore, using the lower tail of matrix Chernoff inequality (\autoref{lem:matrix-chernoff}) we have 
    \begin{align}
        \label{eq:prob-ub-basis-vector}
        \Pr(\rank(G_j^f)< R) = \Pr_f(\lambda_{\min}(G_j(f)) \leq 0) \leq R\cdot e^{-\mu_{\min}/2R} = R\cdot e^{-(2^n-R)/4R}. 
    \end{align}
    We can then bound the probability that \emph{any} standard basis vector $e^t_j$ is in the row span of $H_d(f)$ by the union bound 
    \begin{align}
        \label{eq:final-estimate}
        \Pr\left(\, \exists\, j \text{ such that } e^t_j \text{ in the row span of }H_d(f)\right) & \leq 2^n\cdot R\cdot e^{-(2^n-R)/4R}, \nonumber \\
        & \leq \frac{2^{2n}}{n^{2c}}\cdot  e^{-(n^{2c}-1)/4}. 
    \end{align}
    Choosing $c> 1/2$ we can see that the failure probability in \autoref{eq:final-estimate} is $o(1)$, thus completing the proof of the theorem.  
\end{proof}

An interesting special case of \autoref{thm:ndeg-random} occurs when we consider $\ndeg(f)$ and $\ndeg(\neg f)$ for a random $f$. By \autoref{thm:ndeg-random}, the probability that, for a random $f$, both $\ndeg(f)$ and $\ndeg(\neg f)$ are \emph{not} at most $n/2+O(\sqrt{n\log n})$ is $o(1)$. We thus obtain the following corollary proving a conjecture of \cite{IJKKDKSWW23}. 
\RdegRandom*

Together with a result of \cite{IJKKDKSWW23} it establishes a tight bound on the rational degree of random Boolean functions up to lower order terms.

\section{Effective hypercube nullstellensatz for many polynomials}
\label{sec:nss}

A recent breakthrough work \cite{KKDWY26} proved an effective version of nullstellensatz on Boolean hypercube.
\NssRdeg*
They conjectured, in view of well-known classical nullstellensatz results (see \cite{jelonek05} and references therein), that the following natural generalization of \autoref{thm:nss-rdeg} to any number of polynomial holds. 
\ConjNssRdeg*

In the following we disprove the conjecture even when $m=3$. Note that when $m=2$ we have \autoref{thm:nss-rdeg}. We will need the following proposition from \cite{KKDWY26}. 
\begin{proposition}[{\cite{KKDWY26}}]
    \label{prop:counter-eg-rdeg-partial}
    Let $f_1(x_1,\ldots ,x_n,y_1,\ldots ,y_n) := \sum_{i=1}^nx_i$ and $Z_1 := \{ (a,b) \in \zone^n\times \zone^n \mid f_1(a,b) = 0\}$. Let $f_2(x_1,\ldots ,x_n,y_1,\ldots ,y_n):=\sum_{i=1}^nx_i+ \sum_{j=1}^ny_j-(n+1)$ and $Z_2 := \{ (a,b) \in \zone^n\times \zone^n \mid f_2(a,b) = 0\}$. Further, define $f\colon Z_1 \sqcup Z_2 \to \zone$ by $f(a,b)=0$ if and only if $(a,b)\in Z_1$. Then, 
    any polynomial $p(x_1,\ldots ,x_n,y_1,\ldots ,y_n)$ satisfying $p(a,b)=f(a,b)$ for all $(a,b)\in Z_1\sqcup Z_2$ must have $\deg(p) \geq n$. 
\end{proposition}
We are now ready to present a counterexample to \autoref{conj:nss-rdeg}. 
\CounterNssRdeg*
\begin{proof}
    Let $g_1:=f_1(x,y)\cdot (x_1+x_2)$, $g_2:=f_1(x,y)\cdot (x_1+x_2-1)$, and $g_3:=f_2(x,y)\cdot (x_1+x_2-2)$, where $f_1$ and $f_2$ are as defined in \autoref{prop:counter-eg-rdeg-partial}. Clearly they are non-zero polynomials of degree $2$. Furthermore, it is easily seen that a common zero of $g_1$, $g_2$ and $g_3$ will also be a common zero of $f_1$ and $f_2$. But we know from \autoref{prop:counter-eg-rdeg-partial} that $f_1$ and $f_2$ do not share any common zero, and therefore, $g_1$, $g_2$, and $g_3$ do not share any common zero over the Boolean hypercube. Finally, $g_1(a,b)g_2(a,b)g_3(a,b)=0$ for all $(a,b)\in\zone^n\times\zone^n$ since $(x_1+x_2)(x_1+x_2-1)(x_1+x_2-2)\equiv 0$ over the Boolean hypercube. 

    Let $p_1$, $p_2$, and $p_3$ be any polynomials as given in the statement of \autoref{thm:counter-nss-rdeg}. Further, let $Z_1$ and $Z_2$ be the zero set of $f_1$ and $f_2$, respectively. By definition, $p_3$ evaluates to $0$ on $Z_2$ and since $p_1$ and $p_2$ evaluates to $0$ on $Z_1$, $p_3$ must evaluate to $1$ on $Z_1$. Thus, by \autoref{prop:counter-eg-rdeg-partial}, $\deg(p_3) \geq n$. 
\end{proof}

We also present a counterexample with linear polynomials but the number of polynomials increases to $5$. In fact, \autoref{thm:counter-nss-rdeg} is optimal with respect to the number of polynomials and their degrees, as we show in \autoref{thm:optimal-eg}. 
\begin{proposition}
    \label{prop:counter-nss-rdeg-linear}
    Consider the following non-zero polynomials: $g_1(x,y) := x_1$, $g_2(x,y) := x_2$, $g_3(x,y) := x_1 - x_2$, $g_4(x,y) := f_1(x,y)$, and $g_5(x,y) := f_2(x,y)$, where $f_1$ and $f_2$ are as defined in \autoref{prop:counter-eg-rdeg-partial}. Clearly, they do not share any common zeroes and $\prod_i g_i \equiv 0$ on the Boolean hypercube. 
    Let $p_1,p_2,p_3,p_4,p_5\in\reals[x_1,\ldots x_n,y_1,\ldots ,y_n]$ be any (multilinear) polynomials satisfying  
    \begin{itemize}
        \item $g_i(a,b) = 0 \implies p_i(a,b)=0$ for all $i\in[5]$ and $(a,b)\in\zone^n\times\zone^n$, and 
        \item $p_1(a,b) + p_2(a,b) + p_3(a,b) + p_4(a,b) + p_5(a,b)=1$ for all $(a,b)\in\zone^n\times\zone^n$.
    \end{itemize}
    Then, \[\max\{ \deg(p_1), \deg(p_2), \deg(p_3), \deg(p_4), \deg(p_5) \} \geq n.\]
\end{proposition}
\begin{proof}
    Let $Z_1$ and $Z_2$ be the zero set of $f_1$ and $f_2$, respectively. Note that $p_1$, $p_2$, $p_3$, and $p_4$ all evaluate to $0$ on $Z_1$, and thus $p_5$ must evaluate to $1$ on $Z_1$, while it evaluates to $0$ on $Z_2$ by definition. Hence, by \autoref{prop:counter-eg-rdeg-partial}, $\deg(p_5) \geq n$. 
\end{proof}

Note that an important requirement in nullstellensatz is that polynomials do not share any common zeroes. However, this does not forbid some of them from sharing a common zero. In the following we identify a constraint on the zero sets of polynomials that allows to prove an effective version for many polynomials. 
\NssPartition*

\begin{proof}
    Since $S$ and $[m]\setminus S$ are non-empty, define the following polynomials: 
    \[g_1^{\ast}= \prod_{i\in S}g_i \text{ and } g_2^{\ast} = \prod_{i\notin S}g_i. \]
    Then, by property (iii), we have that $g_1^{\ast}$ and $g_2^{\ast}$ do not share any common zeroes. Furthermore, $g_1^{\ast}(a)\cdot g_2^{\ast}(a) = \prod_{i=1}^{m}g_i(a) =0$ for all $a\in\zone^n$. Therefore, by \autoref{thm:nss-rdeg}, there exists $p_1^{\ast}$ and $p_2^{\ast}$ such that \begin{alphaenumerate}
        \item $g_i^{\ast}(a) = 0 \implies p_i^{\ast}(a)=0$ for all $i\in\{1,2\}$ and $a\in\zone^n$,
        \item $p_1^{\ast}(a)+ p_2^{\ast}(a)=1$ for all $a\in\zone^n$, and 
        \item \(\max\{\deg(p_1^{\ast}), \deg(p_2^{\ast})\} \leq \poly(\deg(g_1^{\ast}),\deg(g_2^{\ast}))\). 
    \end{alphaenumerate} 
    Now fix a $k\in S$ and $j \notin S$. Define $p_i := 0$ for all $i\in[m]\setminus \{k,j\}$, $p_k := p_1^{\ast}$, and $p_j := p_2^{\ast}$. 
    These polynomials satisfy the conclusion of the theorem statement.  
\end{proof}

\subsection{Optimality of the example in \autoref{thm:counter-nss-rdeg}}
\label{sec:optimal-eg}
In this subsection, we show that a counterexample to \autoref{conj:nss-rdeg} with $3$ polynomials, say $g_1$, $g_2$ and $g_3$ must satisfy $\max\{ \deg(g_1),\deg(g_2),\deg(g_3)\}$ at least $2$. 
\OptimalEg* 

\begin{proof}
    Let $g_1,g_2,g_3\colon\zone^n\to\reals$ be non-zero affine-linear forms such that 
    \begin{romanenumerate}
        \item\label{item:covering} $g_1(a)g_2(a)g_3(a)=0$ for all $a\in\zone^n$, and 
        \item\label{item:disjoint} they share no common zeroes in $\zone^n$. 
    \end{romanenumerate}
    For $i\in\{1,2,3\}$, define $Z_i := \{a \in \zone^n \mid g_i(a)=0\}$ and $\eta_i := |Z_i|/2^n$. Thus, by property~(\ref{item:covering}), we have $\eta_1 + \eta_2 +\eta_3 \geq 1$.  We now have two cases based on $g_1,g_2,g_3$. 
    \begin{description}
        \item[Case 1:] Two of the hyperplanes, say $Z_1$ and $Z_2$, cover the full hypercube $\zone^n$ on their own. That is, $\zone^n = Z_1 \cup Z_2$. Then, by \autoref{lem:linear-forms-covering-cube}, we have exactly one of the following subcases. 
        \begin{alphaenumerate}
            \item For some $i$, $g_1=x_i$ and $g_2=1-x_i$. In this case, it suffices to take $h_1=1$, $h_2=1$ and $h_3=0$ to obtain the conclusion in the theorem. 
            \item For some distinct $i,j$, $g_1=x_i-x_j$ and $g_2=x_i+x_j-1$. In such a case, let $h_1=g_1$, $h_2=g_2$ and $h_3=0$.   
        \end{alphaenumerate}
        In both cases we have $\max\{\deg(h_1), \deg(h_2), \deg(h_3)\} \leq 1$. 
         
        \item[Case 2:] No two hyperplanes can cover the full hypercube $\zone^n$. That is, for distinct $i,j\in\{1,2,3\}$, $Z_i\cup Z_j \subset \zone^n$.  Then, we claim that for each $i\in\{1,2,3\}$, $\eta_i \geq 1/4$. 

        Let us complete the proof of theorem assuming the claim. For $i\in\{1,2,3\}$, let $k_i$ be the number of variables with non-zero coefficients in the affine-linear form $g_i$. Then by the Littlewood-Offord inequality (\autoref{lem:littlewood-offord}), we have $\eta_i \leq \sqrt{\frac{2}{\pi k_i}}$, which in turn implies $k_i \leq \frac{2}{\pi \eta_i^2} \leq \frac{32}{\pi} < 11$. Therefore, together $g_i$'s depend on $k_1+k_2+k_3 = O(1)$ variables. Hence the solutions $h_i$'s have degree at most $O(1)$.    

        We are now left to prove the claim that $\eta_i\geq 1/4$ for all $i$. 
        We show it for $\eta_1$. The argument is analogous for others. 
        Consider the function $g'(x) := g_2(x)g_3(x)$ for all $x\in\zone^n$. Since $Z_2\cup Z_3\subset \zone^n$, $g'$ is a non-zero multilinear polynomial of degree at most $2$. Therefore, by \autoref{lem:schwartz-zippel}, $\Pr[g'(x)\neq 0] \geq 1/4$. It is then implied by  property~(\ref{item:covering}) that $\eta_1=\Pr[g_1(x)=0]\geq 1/4$.   
    \end{description}
\end{proof}

We are now ready to characterize a pair of affine-linear forms that together covers the Boolean hypercube. 

\begin{lemma}
    \label{lem:linear-forms-covering-cube}
    Let $A(x)=a_0+\sum_{i=1}^na_ix_i$ and $B(x)=b_0+\sum_{i=1}^nb_ix_i$ be non-zero affine-linear forms over $\reals$. Further, suppose $\zone^n \subseteq \{A=0\}\cup\{B=0\}$. Then, up to multiplication by non-zero constants, and up to relabelling, exactly one of the following holds: 
    \begin{alphaenumerate}
        \item For some $i\in[n]$, $A(x)=x_i$ and $B(x)=1-x_i$. 
        \item For some distinct $i,j\in[n]$, $A(x)=x_i-x_j$ and $B(x)=x_i+x_j-1$. 
    \end{alphaenumerate}
\end{lemma}
\begin{proof}
    We need the following two technical claims, which we prove afterwards. 
    \begin{claim}
        \label{claim:soln-linear-forms}
        For a non-zero affine-linear form $L(x)$ on $n$ variables, 
        \(\# L := |\{z \in \zone^n\mid L(z) = 0\}| \leq 2^{n-1}.\)  
    \end{claim}
    \begin{claim}
        \label{claim:linear-forms-extremal}
        If a non-zero affine-linear form $L(x)$ on $n$ variables is such that $\# L = 2^{n-1}$, then, up to multiplication by a non-zero constant, $L(x) \in \{x_i, 1- x_i, x_i-x_j, x_i+x_j-1\}$, for distinct $i,j\in [n]$. 
    \end{claim}
    Assuming the claims we complete the proof now. By \autoref{claim:soln-linear-forms} and the assumption $\zone^n \subseteq \{A=0\}\cup\{B=0\}$, we must have $\#A =\#B = 2^{n-1}$ and $A\cap B = \emptyset$. By \autoref{claim:linear-forms-extremal}, $A,B\in\{x_i, 1- x_i, x_i-x_j, x_i+x_j-1\}$ and moreover they should be complementary. Therefore, either 
    $A=x_i$ and $B=1-x_i$, or $A=x_i-x_j$ and $B=x_i+x_j-1$. 
\end{proof}
We now prove the two claims. 
\begin{claimproof}[Proof of \autoref{claim:soln-linear-forms}]
    Let $L(x) = a_0 + \sum_{i=1}^na_ix_i$. Since $L$ is non-zero either there exists a variable $x_k$ with the non-zero coefficient $a_k$, or $L$ is a non-zero constant function. If $L$ is a non-zero constant function, then clearly $\#L = 0 \leq 2^{n-1}$, for all $n\geq 0$. Henceforth, we assume, without loss of generality, that the variable $x_n$ has non-zero coefficient.  Thus, for some affine-linear form $A$ on $n-1$ variables, we have 
    \[L(x) = a_nx_n + A(x_1,\ldots ,x_{n-1}).\]
    Now observe that for every setting $y=(x_1,\ldots ,x_{n-1})\in\zone^{n-1}$, there is at most one solution to $x_n$ that satisfies $L(x)=a_nx_n+A(y)=0$. Therefore, the number of solutions to $L$ is at most $2^{n-1}$.

\end{claimproof}
\begin{claimproof}[Proof of \autoref{claim:linear-forms-extremal}]
    Since $L(x)$ is a non-zero affine-linear form, we assume, without loss of generality, up to multiplication by a non-zero constant, 
    \[L(x) = x_i - g(x_1,\ldots ,x_{i-1},x_{i+1},\ldots ,x_n),\]
    for some affine-linear form $g$ in the remaining $n-1$ variables. 
    For every assignment $z\in\zone^{n-1}$ to the variables of $g$, at most one value of $x_i$ can staisfy $L(x)=0$. Since $\#L = 2^{n-1}$, there must be exactly one such zero for each $z$. Therefore, $g(z) \in \zone$ for all $z\in\zone^{n-1}$. Thus, $g$ is a Boolean function of degree at most $1$. That is, $g \in \{0,1,x_j,1-x_j\}$ for $j\neq i$. This completes the proof of the claim.   
\end{claimproof}

\section{Conclusion}
\label{sec:conclusion}
In this paper, we studied non-deterministic polynomial representation of Boolean functions. We showed that almost all Boolean functions have non-deterministic degree as well as rational degree at most $n/2+O(\sqrt{n\log n})$. This  confirms a conjecture of \cite{IJKKDKSWW23} and together with their lower bound establishes tight extremal bounds for random Boolean functions up to lower order terms.    

We also studied effective versions of hypercube nullstellensatz and disproved \autoref{conj:nss-rdeg} of \cite{KKDWY26}. On the positive side, we proved a variant of \autoref{conj:nss-rdeg}. We also show the optimality of our counterexample to \autoref{conj:nss-rdeg}. 

For the future work, one of the most interesting question that remains open is to close the gap between $\deg$/$\dqc$ and $\ndeg$/$\rdeg$. In particular, de Wolf \cite{deWolf00} conjectured that \(\dqc(f) = O(\ndeg(f)\ndeg(\neg f))\). The current best known separation is quadratic (see \cite{KKDWY26}). Even the weaker versions of this conjecture, namely $Q_E(f)= O(\ndeg(f)\ndeg(\neg f))$ and $\deg(f)= O(\ndeg(f)\ndeg(\neg f))$, remains open. 

The $\deg(f)\overset{?}{=} O(\ndeg(f)\ndeg(\neg f))$ question is also interesting from the perspective of effective hypercube nullstellensatz. If true it would match the well-known effective version of classical nullstellensatz \cite{jelonek05}.

\bibliography{ndeg-bib}

\end{document}